\documentclass[3p,times]{elsarticle}

\usepackage{ecrc}

\volume{00}

\firstpage{1}

\journalname{Nuclear Physics A}

\runauth{L. Tolos et al.}

\jid{NUPHA}

\jnltitlelogo{Nuclear Physics A}

\CopyrightLine{2012}{Published by Elsevier Ltd.}

\usepackage{amssymb}

\usepackage[figuresright]{rotating}

\begin{document}

\begin{frontmatter}

\dochead{}



\title{Strangeness and Charm in Nuclear Matter}

\author[a,b]{Laura Tolos}\author[c]{Daniel Cabrera}\author[d]{Carmen Garcia-Recio}\author[e]{Raquel Molina}\author[f]{Juan Nieves}\author[f]{Eulogio Oset}\author[g]{Angels Ramos}\author[h]{Olena Romanets}\author[d]{Lorenzo Luis Salcedo}

\address[a]{ Instituto de Ciencias del Espacio (IEEC/CSIC), Campus Universitat 
Aut\`onoma de Barcelona, Facultat de Ci\`encies, Torre C5, E-08193 Bellaterra 
(Barcelona), Spain}
\address[b]{Frankfurt Institute for Advanced Studies, Johann Wolfgang Goethe University, Ruth-Moufang-Str. 1,
60438 Frankfurt am Main, Germany}
\address[c]{Departamento de F\'{\i}sica Te\'orica II, Universidad Complutense,
28040 Madrid, Spain}
\address[d]{Departamento de F{\'\i}sica At\'omica, Molecular y Nuclear, and Instituto Carlos I de F{\'i}sica Te\'orica y Computacional,
Universidad de Granada, E-18071 Granada, Spain}
\address[e]{Research Center for Nuclear Physics (RCNP),
Mihogaoka 10-1, Ibaraki 567-0047, Japan}
\address[f]{Instituto de F{\'\i}sica Corpuscular (centro mixto CSIC-UV),
Institutos de Investigaci\'on de Paterna, Aptdo. 22085, 46071, Valencia, Spain}
\address[g]{Departament d'Estructura i Constituents de la Mat\`eria,
Universitat de Barcelona,
Diagonal 647, 08028 Barcelona, Spain}
\address[h]{Theory Group, KVI, University of Groningen,
Zernikelaan 25, 9747 AA Groningen, The Netherlands}

\begin{abstract}
The properties of strange ($K$, $\bar K$ and $\bar K^*$) and open-charm ($D$, $\bar D$ and $D^*$) mesons in dense matter are studied using a unitary approach in coupled channels for meson-baryon scattering. In the strangeness sector, the interaction with nucleons always comes through vector-meson exchange, which is evaluated by chiral  and hidden gauge Lagrangians.
For the interaction of charmed mesons with nucleons we extend the SU(3) Weinberg-Tomozawa Lagrangian to incorporate spin-flavor symmetry and implement a suitable flavor symmetry breaking. The in-medium solution for the scattering amplitude  accounts for Pauli blocking effects and meson self-energies. On one hand,  we obtain the $K$, $\bar K$ and $\bar K^*$ spectral functions in the nuclear medium and study their behaviour at finite density, temperature and momentum.  We also make an estimate of  the transparency ratio of the $\gamma A \to K^+ K^{*-} A^\prime$ reaction, which we propose as a tool to detect  in-medium modifications of the $\bar K^*$ meson. On the other hand, in the charm sector, several resonances with negative parity are generated dynamically by the $s$-wave interaction between pseudoscalar and vector meson multiplets with $1/2^+$ and $3/2^+$ baryons. The properties of these states in matter are analyzed and their influence on the open-charm meson spectral functions is studied. We finally discuss  the possible formation of $D$-mesic nuclei at FAIR energies.

\end{abstract}

\begin{keyword}
Strange mesons \sep transparency ratio  \sep dynamically-generated baryon resonances \sep open charm in matter \sep $D$-mesic nuclei

\end{keyword}

\end{frontmatter}


\section{Introduction}
\label{intro}

Strangeness and charm in hot and dense matter is a matter of extensive analysis in connection to heavy-ion collisions from SIS \cite{Fuchs:2005zg} to FAIR \cite{fair} energies at GSI.

In the strange sector, the interaction of strange pseudoscalar mesons ($K$ and
$\bar K$) with matter is a topic of high interest. Whereas the interaction of
$\bar K N$ is repulsive at threshold, the phenomenology of antikaonic atoms
\cite{Friedman:2007zz} shows that the $\bar K$ feels an attractive potential at
low densities. This attraction is a consequence of the modified $s$-wave
$\Lambda(1405)$ resonance in the medium due to Pauli blocking effects
\cite{Koch} together with the self-consistent consideration of the $\bar K$
self-energy \cite{Lutz} and the inclusion of self-energies of the mesons and
baryons in the intermediate states \cite{Ramos:1999ku}. Attraction of the order
of -50 MeV at normal nuclear matter density, $\rho_0=0.17 \,{\rm fm^{-3}}$, is
obtained in unitarizated theories in coupled channels based on chiral dynamics
\cite{Ramos:1999ku} and meson-exchange models \cite{Tolos01,Tolos02}. Moreover,
the knowledge of higher-partial waves beyond $s$-wave
\cite{Tolos:2006ny,Lutz:2007bh,Tolos:2008di} becomes essential for relativistic
heavy-ion experiments at beam energies below 2AGeV \cite{Fuchs:2005zg}.
 As for vector mesons in the nuclear medium, only the
non-strange ones have been the main focus of attention for years, while 
very little discussion has been made about the properties of the strange ones,
$K^*$ and $\bar K^*$.  Only recently the $\bar
K^*N$ interaction in free space has been addressed in
Ref.~\cite{GarciaRecio:2005hy} using SU(6) spin-flavour symmetry, and in
Refs.~\cite{Oset:2009vf, Khemchandani:2011et} within the hidden local gauge
formalism for the interaction of vector mesons with baryons of the octet and the
decuplet. Within the scheme of Ref.~\cite{Oset:2009vf}, medium effects have been
implemented and analyzed very recently \cite{tolos10} finding an spectacular
enhancement of the $\bar{K}^*$ width in the medium.

With regard to the charm sector, the nature of new charmed and strange hadron resonances is an active topic of research, with recent data coming from  CLEO, Belle, BaBar  and other experiments \cite{facility00}. In the upcoming years, the experimental program of the future FAIR facility at GSI \cite{fair}  will also face new challenges where charm plays a dominant role.  Approaches based on coupled-channel dynamics have proven to be very successful in describing the experimental data. In particular, unitarized coupled-channel methods have been applied in the meson-baryon sector with charm content \cite{Tolos:2004yg,Tolos:2005ft,Lutz:2003jw,Hofmann:2005sw,Mizutani:2006vq,Tolos:2007vh,JimenezTejero:2009vq}, partially motivated by the parallelism between the $\Lambda(1405)$ and the $\Lambda_c(2595)$. Other existing coupled-channel approaches are based on the J\"ulich meson-exchange model \cite{Haidenbauer:2007jq,Haidenbauer:2010ch} or on the hidden gauge formalism \cite{Wu:2010jy}.

However, these models are not fully consistent with heavy-quark spin symmetry (HQSS) ~\cite{Isgur:1989vq}, which is a proper QCD symmetry that appears when the quark masses, such as the charm mass, become larger than the typical confinement scale. Aiming at incorporating HQSS, an SU(6) $\times$ SU(2) spin-flavor symmetric model has been recently developed \cite{GarciaRecio:2008dp,Gamermann:2010zz}, similarly to the SU(6) approach in the light sector of Refs.~\cite{GarciaRecio:2005hy,Toki:2007ab}. The model generates dynamically resonances with negative parity in all the isospin, spin, strange and charm sectors  that one can form from an s-wave interaction between pseudoscalar and vector meson multiplets with $1/2^+$ and $3/2^+$ baryons \cite{Romanets:2012hm}. Recent calculations on bottomed resonances using this HQSS model  have also been performed \cite{GarciaRecio:2012db}.

In this paper we review the properties of the strange ($K$, $\bar K$ and $\bar K^*$)  and open-charm ($D$, $\bar D$ and $D^*$) mesons in dense matter. We address different experimental scenarios that can be analyzed in present and future experiments so as to test the properties of these mesons in matter, such as the transparency ratio of the $\gamma A \to K^+ K^{*-} A^\prime$ reaction and the formation of $D$-mesic nuclei.

\section{$K$ and $\bar K$ mesons in a hot dense nuclear medium}

The  kaon and antikaon self-energies in symmetric nuclear matter at finite
temperature are obtained from the $s$- and $p$-waves in-medium kaon-nucleon
interaction within a chiral unitary approach \cite{Tolos:2008di,ewsr}. The
$s$-wave amplitude of the $\bar K N$ comes, at tree level, from the
Weinberg-Tomozawa term of the chiral Lagrangian. Unitarization in coupled
channels is imposed on on-shell amplitudes ($T$) with a cutoff regularization.
The $\Lambda(1405)$ resonance in the $I=0$ channel is generated dynamically and
 a satisfactory description of low-energy scattering observables is
achieved. The $K N$ effective interaction is also obtained from the
Bethe-Salpeter equation using the same cutoff parameter. 

 The in-medium solution of the $s$-wave amplitude accounts for Pauli-blocking effects, mean-field binding on the nucleons and hyperons via a $\sigma-\omega$ model, and the dressing of the pion and kaon propagators. The self-energy is then obtained in a self-consistent manner summing the transition amplitude $T$ for the different isospins over the nucleon Fermi distribution at a given temperature, $n(\vec{q},T)$,  
\begin{eqnarray}
\Pi_{K(\bar K) N}(q_0,{\vec q},T)= \int \frac{d^3p}{(2\pi)^3}\, n(\vec{p},T) \,
[\, {T}_{K(\bar K)N}^{(I=0)} (P_0,\vec{P},T) +
3 \, {T}_{K(\bar K)N}^{(I=1)} (P_0,\vec{P},T)\, ], \ \  \ \ \ \label{eq:selfd}
\end{eqnarray}
where $P_0=q_0+E_N(\vec{p},T)$ and $\vec{P}=\vec{q}+\vec{p}$ are
the total energy and momentum of the kaon-nucleon pair in the nuclear
matter rest frame, and ($q_0$, $\vec{q}\,$) and ($E_N$, $\vec{p}$\,) stand  for
the energy and momentum of the kaon and nucleon, respectively, also in this
frame. In the case of  the  $\bar K$ meson the model includes, in
addition, a $p$-wave contribution to the self-energy from hyperon-hole ($Yh$)
excitations, where $Y$ stands for $\Lambda$, $\Sigma$ and
$\Sigma^*$ components. For the $K$ meson the $p$-wave self-energy results from
$YN^{-1}$ excitations in crossed kinematics. The self-energy determines,
through the Dyson equation, the in-medium kaon propagator and the corresponding
kaon spectral function.

\begin{figure}[htb]
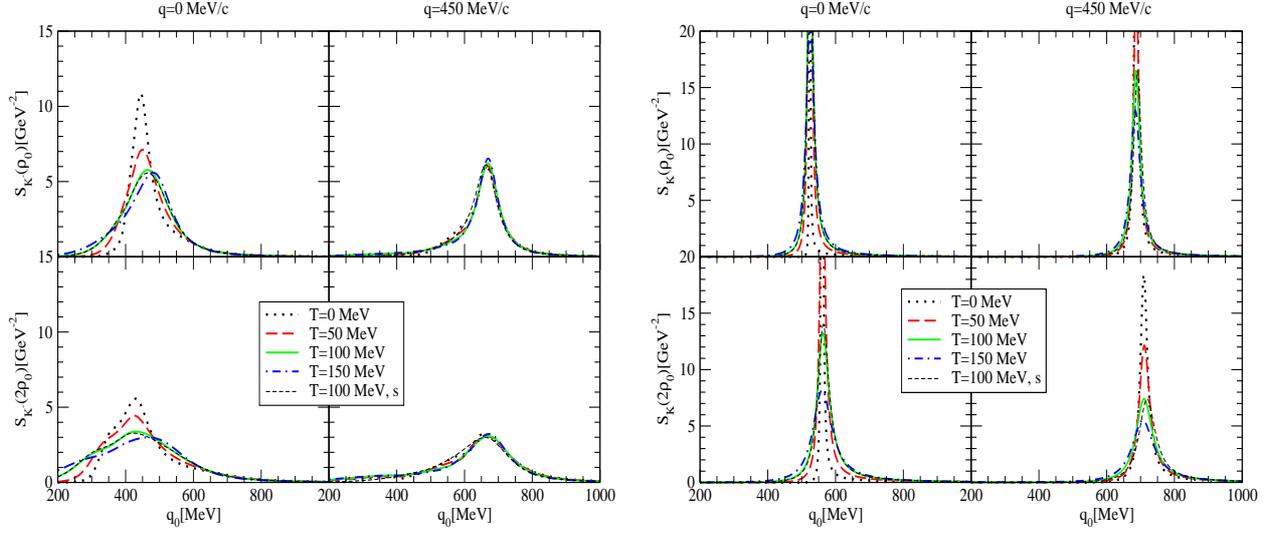

\begin{center}
\includegraphics[height=7 cm, width=8cm]{fig6.eps}
\hfill
\includegraphics[height=7 cm, width=8cm]{fig7.eps}
\caption{ $\bar K$ (left) and $K$ (right) spectral functions for different densities, temperatures and momenta.}
 \label{fig1}
\end{center}
\end{figure}

The evolution of the  $\bar{K}$ and $K$ spectral functions with density and
temperature is shown in Fig.~\ref{fig1}. The $\bar K$ spectral function
(left) shows a broad peak that results from a strong mixing between the
quasi-particle peak and the $\Lambda(1405)N^{-1}$ and
$Y(=\Lambda, \Sigma , \Sigma^*)N^{-1}$ excitations. 
These
$p$-wave $YN^{-1}$ subthreshold excitations affect mainly the properties of
the ${\bar K}$ at finite momentum, enhancing the low-energy tail of the spectral
function and providing a
repulsive contribution to the $\bar K$ potential that
partly compensates the attraction obtained from the stronger $s$-wave ${\bar K}
N$ interaction component. Temperature and density soften the $p$-wave
contributions to the spectral function at the quasi-particle energy. As for the
$K$ meson, the spectral function (right) shows a narrow
quasi-particle peak which dilutes with temperature and density as the phase
space for $KN$ states increases. With increasing density, the repulsive
character of the $s$-wave $K N$ interaction also increases, thereby shifting
the quasi-particle peak of the $K$ spectral function to higher energies.

There has been a lot of activity aiming at extracting the properties of kaons
and antikaons in a dense and hot environment from heavy-ion collisions.
Some time ago, the antikaon production in nucleus-nucleus collisions at SIS
energies was studied using a BUU transport model with antikaons that were
dressed with the Juelich meson-exchange model \cite{Cassing:2003vz}.
Multiplicity ratios involving strange mesons coming from heavy-ion
collisions data were also analyzed \cite{Tolos:2003qj}. More recently,
a systematic study of the experimental results of KaoS collaboration
was performed together with a detailed comparison to transport model
calculations \cite{Forster:2007qk}. Several conclusions on the production
mechanisms for strangeness were achieved. However, to which extent the
properties of $\bar K$ mesons are modified in matter is not fully
understood. There is still no convincing
simultaneous description of all experimental data that involve antikaons in
matter. These conclusions can also be found in a recent report on
strangeness production close to threshold in proton-nucleus and heavy-ion
collisions \cite{Hartnack:2011cn}.

\begin{figure}[t]
\begin{center}
\includegraphics[width=0.45\textwidth,height=6cm]{spectral_ksn.eps}
\hfill
\includegraphics[width=0.5\textwidth,height=6.5cm]{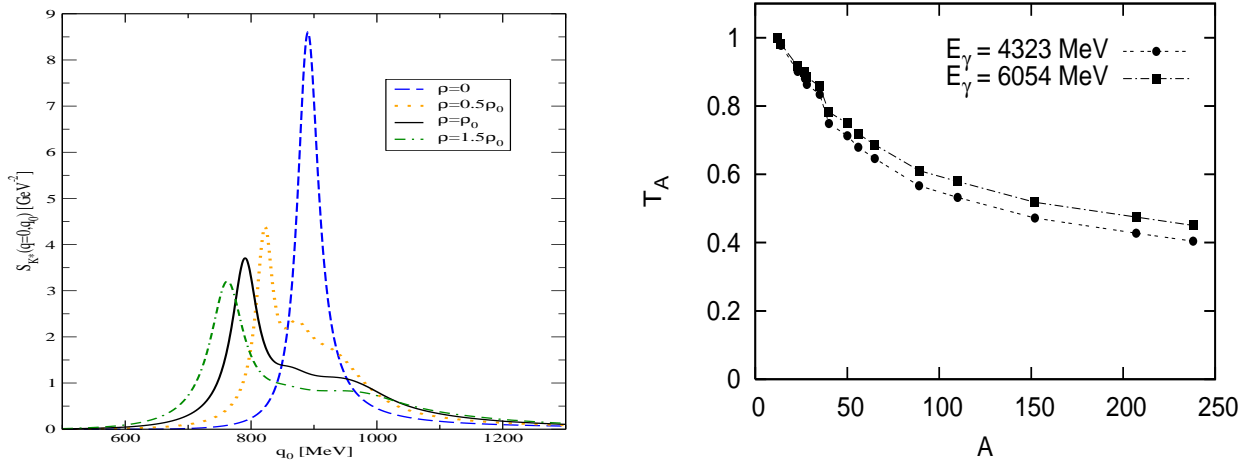}
\caption{ Left: $\bar K^*$ spectral function at $\vec{q}=0$ MeV/c for different densities. Right: Transparency ratio for  $\gamma A \to K^+ K^{*-} A'$}
\label{fig:spec-trans}
\end{center}
\end{figure}

\section{$\bar K^*$ meson in matter}

The  $\bar K^*$  self-energy in symmetric nuclear matter is obtained  within the
hidden gauge formalism \cite{tolos10}. There are two sources for the
modification of the $\bar K^*$ $s$-wave self-energy in nuclear matter: a) the
contribution associated to the decay mode $\bar K \pi$ modified by nuclear
medium effects on the $\bar K$ and $\pi$ mesons,  which accounts for
the $\bar K^* N \to \bar K N, \pi Y, \bar K \pi N, \pi \pi Y \dots$ processes, 
with $Y=\Lambda,\Sigma$, and b) the contribution associated to the interaction
of the $\bar K^*$ with the nucleons in the medium, which accounts for the direct
quasi-elastic process $\bar K^* N \to \bar K^* N$, as well as other absorption
channels involving vector mesons, $\bar K^* N\to \rho Y, \omega Y, \phi Y,
\dots$. In fact, this last term comes from a unitarized coupled-channel
process, similar to the $\bar K N$ case. Two resonances are generated
dynamically, $\Lambda(1783)$ and $\Sigma(1830)$, which can be identified with
the experimentally observed states $J^P=1/2^-$ $\Lambda(1800)$ and the
$J^P=1/2^-$ PDG state $\Sigma(1750)$, respectively \cite{Oset:2009vf}.

The in-medium $\bar K^*$ self-energy results from the sum of both contributions,
$\Pi_{\bar K^*}=\Pi_{\bar{K}^*}^{{\rm (a)}}
+\Pi_{\bar{K}^*}^{{\rm (b)}}$,
 where $\Pi_{\bar{K}^*}^{{\rm (b)}}$ is obtained, similarly to the $\bar K$
meson in Eq.~(\ref{eq:selfd}), by integrating the $\bar K^* N$ transition
amplitude over the nucleon Fermi sea.
%

The $\bar K^*$ meson spectral function, which results from the imaginary part of
the in-medium $\bar K^*$ propagator, is displayed in the left panel of
Fig.~\ref{fig:spec-trans} as a function of the meson energy $q_0$, for zero
momentum and different densities up to 1.5 $\rho_0$. The dashed line refers to
the calculation in free space, where only the $\bar K \pi$ decay channel
contributes, while the other three lines correspond to the fully self-consistent
calculations,  which incorporate the process $\bar K^* \rightarrow \bar K
\pi$ in the medium, as well as the quasielastic $\bar K^* N \to \bar K^* N$ and
other $\bar K^* N\to V B$ processes.  The structures above the quasiparticle
peak correspond to the dynamically generated $\Lambda(1783) N^{-1}$ and
$\Sigma(1830) N^{-1}$ excitations. Density effects result in a dilution and
merging of those resonant-hole states, together with a general broadening of the
spectral function  due to the increase of collisional and absorption processes.
Although the real part of the optical potential is moderate, -50 MeV at
$\rho_0$, the interferences with the resonant-hole modes push the $\bar{K}^*$
quasiparticle peak to lower energies. However, transitions to
meson-baryon states with a pseudoscalar meson, which are included in the
$\bar K^*$ self-energy but are not incorporated explicitly in the unitarized
$\bar K^* N$ amplitude, would make the peak less prominent and difficult to
disentangle from the other excitations. In any case, what is clear from the
present approach, is that the spectral function spread of the $\bar K^*$
increases substantially in the medium, becoming at normal nuclear matter density
five times bigger than in free space.

\subsection{Transparency ratio for  $\gamma A \to K^+ K^{*-} A'$}

The nuclear transparency ratio can be studied in order to test
experimentally the $\bar K^*$ self-energy. The idea is to compare the cross
sections of the photoproduction reaction $\gamma A \to K^+ K^{*-} A'$ in
different nuclei, and trace them to the in medium $K^{*-}$ width.

The normalized nuclear transparency ratio is defined as
\begin{equation}
T_{A} = \frac{\tilde{T}_{A}}{\tilde{T}_{^{12}C}} \hspace{1cm} ,{\rm with} \ \tilde{T}_{A} = \frac{\sigma_{\gamma A \to K^+ ~K^{*-}~ A'}}{A \,\sigma_{\gamma N \to K^+ ~K^{*-}~N}} \ .
\end{equation}
 It describes the loss of flux of $K^{*-}$ mesons in the nucleus and is related to the absorptive part of the $K^{*-}$-nucleus optical potential and, thus, to the $K^{*-}$ width in the nuclear medium.  We evaluate the ratio between the nuclear cross sections in heavy nuclei and a light one ($^{12}$C), $T_A$, so that other nuclear effects not related to the absorption of the $K^{*-}$ cancel.

In the right panel of  Fig. \ref{fig:spec-trans} we show the results for
different nuclei. The transparency ratio has been plotted for two different
energies in the center of mass reference system $\sqrt{s}=3$ GeV and $3.5$ GeV,
which are equivalent to energies of the photon in the lab frame of $4.3$ GeV and
$6$ GeV,  respectively. We observe a very strong attenuation of the $\bar{K}^*$
survival probability due to the decay  $\bar{K}^*\to
\bar{K}\pi$ or absorption channels 
$\bar{K}^*N\to \bar K N, \pi Y, \bar K \pi N, \pi \pi Y, \bar K^* N, \rho Y,
\omega Y, \phi Y, \dots$  with increasing nuclear-mass number $A$. This is due
to the larger path that the $\bar{K}^*$ has to follow before it leaves the
nucleus, having then more chances to decay or get absorbed.

\section{Open-charm mesons in nuclear matter with heavy-quark spin symmetry}
\label{medium}

The HQSS predicts that, in QCD, all types of spin interactions involving heavy quarks vanish for infinitely massive quarks. Thus, HQSS connects vector and pseudoscalar mesons containing charmed quarks.  Chiral symmetry fixes  the lowest order interaction between Goldstone bosons and other hadrons  in a model independent way; this is the Weinberg-Tomozawa (WT) interaction. Then, it is very appealing to have a predictive model for four flavors including all basic hadrons (pseudoscalar and vector mesons, and $1/2^+$ and $3/2^+$ baryons) which reduces to the WT interaction in the sector where Goldstone bosons are involved and which incorporates HQSS in the sector where charm quarks participate. Indeed, this is a model assumption which is justified in view of the reasonable semiqualitative outcome of the SU(6) extension in the three-flavor sector \cite{Gamermann:2011mq} and on a formal plausibleness on how the SU(4) WT interaction in the charmed pseudoscalar meson-baryon sector comes out in the vector-meson exchange picture.

The model extension is given schematically by
\begin{equation}
\label{symbolic}
{\mathcal L}^{{\rm SU (8)}}_{ \rm WT} 
=
 \frac{1}{f^2} [[M^{\dagger} \otimes M]_{\bf 63_{a}}
  \otimes [B^{\dagger} \otimes B]_{ \bf 63} ]_{ \bf 1} ,
\end{equation}
which represents the interaction between baryons (in the 120 irrep of SU(6) $\times$ SU(2)) and mesons (in the 63) through $t$-channel exchange in the 63. In the $s$-channel, the meson-baryon space reduces into four SU(6) $\times$ SU(2) irreps, from which two multiplets ${\bf 120}$ and ${\bf 168}$ are the most attractive. As a consequence, dynamically-generated baryon resonances are most likely to occur in those sectors, and therefore we will concentrate on states which belong to these two representations. From this Lagrangian, we can extract the potential model for meson-baryon interaction that respects HQSS \cite{Romanets:2012hm} and solve the on-shell Bethe-Salpeter equation in coupled channels so as to calculate the scattering amplitudes. The poles of the scattering amplitudes are the dynamically-generated baryon resonances. 

Dynamically generated states in different charm and strange sectors  are
predicted within our model
\cite{GarciaRecio:2008dp,Gamermann:2010zz,Romanets:2012hm}. Some of them can be
identified with known states from the PDG \cite{Nakamura:2010zzi}. This
identification is made by comparing the PDG data on these states with the mass,
width and, most important, the coupling to the meson-baryon channels of our
dynamically-generated poles.  In the $C=1,S=0, I=0$ sector, we obtain three
$\Lambda_c$ and one $\Lambda_c^*$ states. The experimental $\Lambda_c(2595)$
resonance can be identified with the pole that we obtain around 2618.8 MeV as
similarly done in Ref.~\cite{GarciaRecio:2008dp}. A second broad $\Lambda_c$
resonance at 2617~MeV is, moreover, observed with a large coupling to the open
channel $\Sigma_c \pi$, very close to  the $\Lambda_c(2595)$. This is the
same two-pole pattern found in the charmless $I = 0, S = -1$ sector for the
$\Lambda(1405)$\cite{Jido:2003cb}. A third spin-$1/2$ $\Lambda_c$ resonance is
seen around 2828~MeV  and cannot be assigned to any experimentally known
resonance. 
With regard to spin-$3/2$ resonances, we find one located at $(2666.6 -i 26.7 \ \rm{MeV})$ that is assigned to the experimental $\Lambda_c(2625)$. For $C=1, S=0, I=1$ ($\Sigma_c$ sector), three $\Sigma_c$ resonances are obtained with masses 2571.5, 2622.7 and 2643.4~MeV and widths 0.8, 188.0 and 87.0~MeV, respectively. Moreover, two spin-$3/2$ $\Sigma_c$ resonances are generated dynamically. The first one, a bound state at 2568.4~MeV, is thought to be the charmed counterpart of the $\Sigma(1670)$. The second state at $2692.9 -i 33.5$ ~MeV has not a direct experimental comparison.

The in-medium modifications of these resonant states will have important consequences on the properties of open-charm mesons in matter. These modifications can be studied by analyzing the self-energy of open-charm in the nuclear medium. The self-energy and, hence, spectral function for $D$ and $D^*$ mesons are obtained self-consistently in a simultaneous manner, as it follows from HQSS, by taking, as bare interaction, the extended WT interaction previously described. We incorporate Pauli blocking effects and open charm meson self-energies in the intermediate propagators for the in-medium solution  \cite{tolos09}. More specifically, the $D$ and $D^*$ self-energies are obtained by summing the
transition amplitude over the Fermi sea of nucleons, as well as over the
different spin ($J=1/2$ for $DN$, and $J=1/2,3/2$ for $D^* N$) and isospin
($I=0,1$) channels.

\begin{figure}
\begin{center}
\includegraphics[width=0.5\textwidth,height=5.8cm]{art_spec.eps}
\hfill
\includegraphics[width=10cm, angle=90,height=6cm]{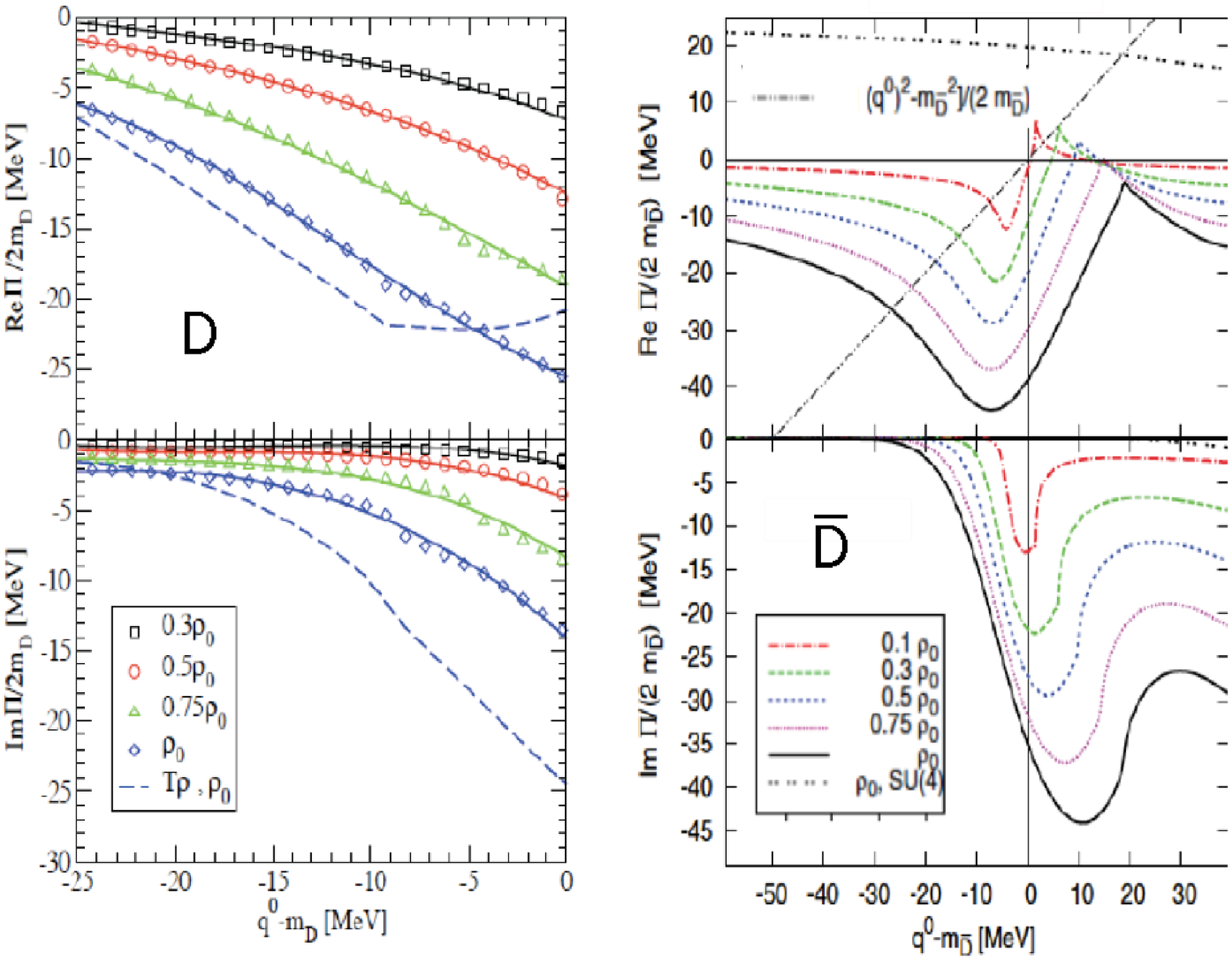}
\caption{Left: The $D$ and $D^*$ spectral functions in dense nuclear matter at $\vec{q}=0$ MeV/c. Right:  The $D$ and $\bar D$ optical potential at $\vec{q}=0$ MeV/c for different densities }
\label{fig2}
\end{center}
\end{figure}

\subsection{D mesic nuclei}

$D$ and $\bar D$-meson bound states in $^{208}$Pb were predicted in Ref.~\cite{tsushima99} relying upon an attractive  $D$ and $\bar D$ -meson potential in the nuclear medium. This potential was obtained within a quark-meson coupling (QMC) model \cite{Guichon:1987jp}. The experimental observation of those bound states, though, might be problematic since, even if there are bound states, their widths could be very large compared to the separation of the levels. This is indeed the case for the potential derived from a SU(4) $t$-vector meson exchange model for $D$-mesons \cite{Tolos:2007vh}.

We solve the Schr\"odinger equation in the local density approximation so as to analyze the formation of bound states with charmed mesons in nucleus. We use the energy dependent optical potential
\begin{equation}
  V(r,E) = \frac{
  \Pi(q^0=m+E,\vec{q}=0,~\rho(r))}{2 m},
\label{eq:UdepE}
\end{equation}
where $E=q^0-m$ is the $D$ or $\bar D$ energy excluding its mass, and $\Pi$ the meson self-energy. The optical potential for different densities is displayed on the r.h.s of Fig.~\ref{fig2}. For $D$ mesons we observe a strong energy dependence of the potential close to the $D$ meson mass due to the mixing of the quasiparticle peak with the  $\Sigma_c(2823)N^{-1}$ and $\Sigma_c(2868)N^{-1}$ states. As for the $\bar D$ meson, the presence of a bound state at 2805 MeV \cite{Gamermann:2010zz}, almost at $\bar D N$ threshold, makes the potential also strongly energy dependent. This is in contrast to the SU(4) model (see Ref.~\cite{carmen10}).

The $D$ and $D^*$ spectral functions are displayed on the l.h.s. of  Fig.~\ref{fig2}. Those spectral functions show a rich spectrum of resonance-hole states. On one hand, the $D$ meson quasiparticle peak mixes strongly with $\Sigma_c(2823)N^{-1}$ and $\Sigma_c(2868)N^{-1}$ states. On the other hand, the $\Lambda_c(2595)N^{-1}$ is clearly visible in the low-energy tail.  With regard to the $D^*$ meson, the $D^*$ spectral function incorporates the $J=3/2$ resonances, and the quasiparticle peak fully mixes with the  $\Sigma_c(2902)N^{-1}$ and $\Lambda_c(2941)N^{-1}$ states.  For both mesons, the $Y_cN^{-1}$ modes tend to smear out and the spectral functions broaden with increasing phase space, as seen before in the ${\rm SU(4)}$ model \cite{Mizutani:2006vq}. Note that resonances with higher masses than those described in Sec.~\ref{medium} are also present in the spectral functions. Those resonant states were seen in the wider energy range explored in Ref.~\cite{GarciaRecio:2008dp} and are not coming from the two most attractive representations, ${\bf 120}$ and ${\bf 168}$.

\begin{figure}[t]
\begin{center}
\includegraphics[width=0.3\textwidth,angle=-90]{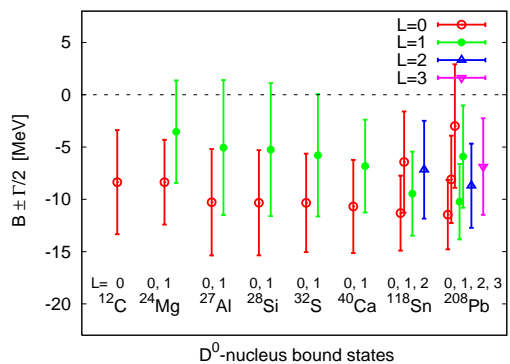}
\caption{$D^0$-nucleus bound states. \label{fig3}}
\end{center}
\end{figure}

Then, the question is whether $D$ and/or $\bar D$  will be bound in nuclei.  We observe that the $D^0$-nucleus states are weakly bound (see Fig.~\ref{fig3}), in contrast to previous results using the QMC model. Moreover,  those states have significant widths \cite{carmen10}, in particular, for $^{208}$Pb \cite{tsushima99}. Only $D^0$-nucleus bound states are possible since the Coulomb interaction prevents the formation of observable bound states for $D^+$ mesons.

\begin{figure}[t]
\begin{center}
\includegraphics[width=0.47\textwidth]{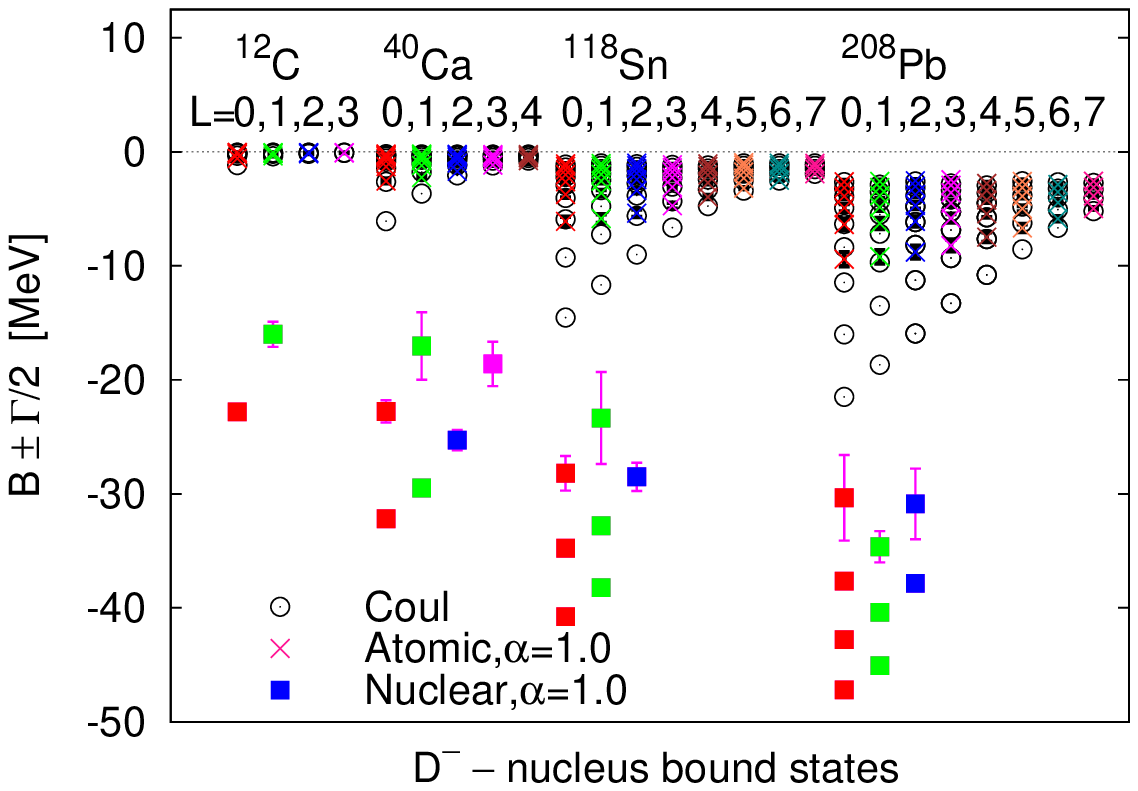}
\hfill
\includegraphics[width=0.47\textwidth]{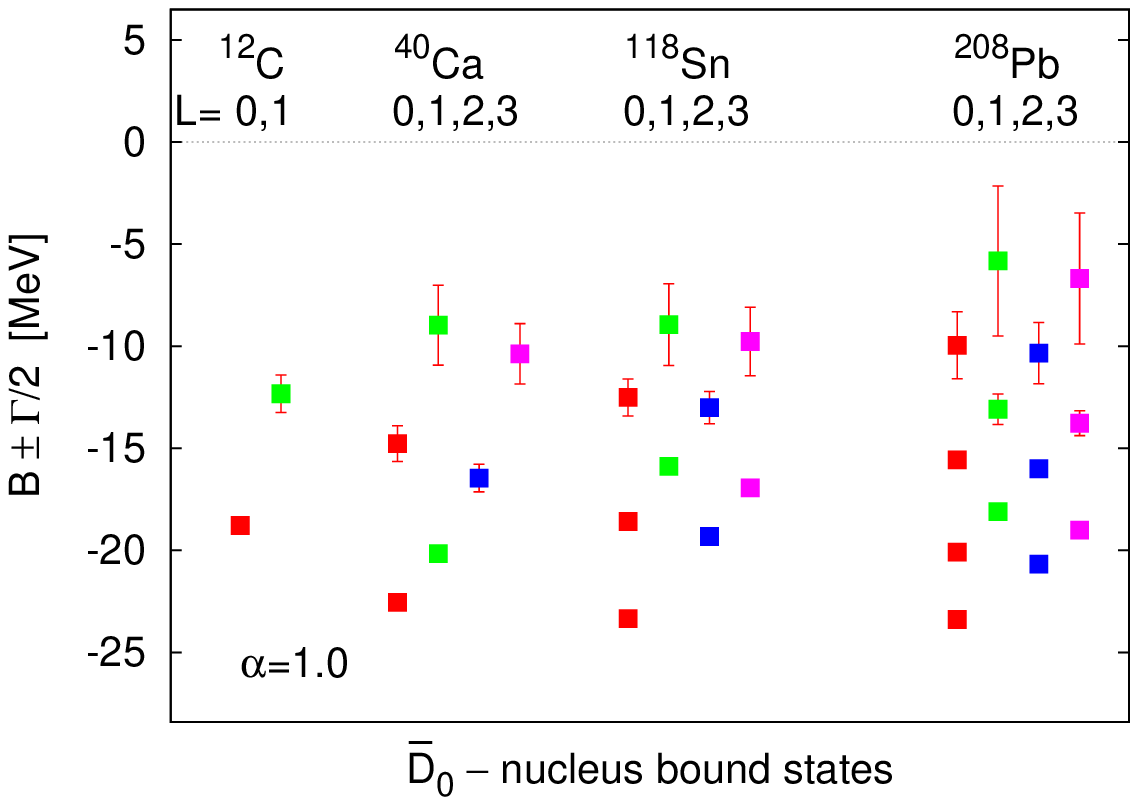}
\caption{$D^-$ and $\bar D^0$- nucleus bound states. \label{fig4}}
\end{center}
\end{figure}

With regard to $\bar D$-mesic nuclei, not only $D^-$ but also $\bar{D}^0$ bind in nuclei (Fig.~\ref{fig4}). The spectrum 
contains states of atomic and of nuclear types for all nuclei for $D^-$  while, as expected, only nuclear states are present for $\bar{D}^0$ in nuclei. Compared to the pure Coulomb levels, the atomic states are less bound. The nuclear ones are more bound and may present a sizable width \cite{GarciaRecio:2011xt}. Moreover, nuclear states only exist for low angular momenta.
It is also interesting to note that a recent work suggests the possibility that, for the lightest nucleus, $DNN$ develops a bound and narrow state with $S=0, I=1/2$, as evaluated in Ref.~\cite{bayar}.

To gain some knowledge on the charmed meson-nucleus interaction, the information on bound states is very valuable. This is of special interest  for PANDA at FAIR. The experimental detection of $D$ and $\bar D$-meson bound states is, however, a difficult task. For example, it was observed in Ref.~\cite{carmen10} that reactions with antiprotons on nuclei for obtaining $D^0$-nucleus states 
might have a very low production rate. Reactions but with proton beams seem more likely to trap a $D^0$ in nuclei \cite{carmen10}. 

\section*{Acknowledgments}

 This research was supported by DGI and FEDER
funds, under Contract Nos. FIS2011-28853-C02-02,
 FIS2011-24149, FIS2011-24154, FPA2010-16963 and the
Spanish Consolider-Ingenio 2010 Programme CPAN
(CSD2007-00042), by Junta de Andaluc\' ia Grant
No. FQM-225, by Generalitat Valenciana under Contract
No. PROMETEO/2009/0090, by the Generalitat de Catalunya under
contract 2009SGR-1289 and by the EU
HadronPhysics3 project, Grant Agreement No. 283286.
O. R.  wishes to acknowledge support from the
Rosalind Franklin Programme. L. T. acknowledges support
from Ramon y Cajal Research Programme, and from FP7-
PEOPLE-2011-CIG under Contract No. PCIG09-GA-
2011-291679.








\end{document}